\documentclass{svproc}

\usepackage{cite}
\usepackage{amsmath,amssymb,amsfonts}
\usepackage{graphicx}
\usepackage{textcomp}
\usepackage{xcolor}
\usepackage{tabularx}
\usepackage{subcaption}

\def\BibTeX{{\rm B\kern-.05em{\sc i\kern-.025em b}\kern-.08em
    T\kern-.1667em\lower.7ex\hbox{E}\kern-.125emX}}
\newcommand{\getsr}{{\:{\leftarrow{\hspace*{-3pt}\raisebox{.75pt}{$\scriptscriptstyle\$$}}}\:}}
\newcommand{\vect}[1]{\mathbf{#1}}

\newcommand{\fname}{\mathsf}
\newcommand{\aname}[1]{\mathbf{\fname{#1}}}
\newcommand{\vname}{\mathsf}
\newcommand{\concat}{\parallel}
\newcommand{\ind}{\approx}

\newcommand{\sind}{\underset{S}{\ind}}
\newcommand{\cind}{\underset{C}{\ind}}
\newcommand{\samplefrom}{\getsr}
\newcommand{\dotprod}[2]{\langle #1, #2 \rangle}

\newcommand{\params}{\vname{params}}
\newcommand{\nblocks}{\vname{numBlocks}}
\newcommand{\records}{\vname{records}}
\newcommand{\ids}{\vname{ids}}
\newcommand{\hscans}{\vname{hscans}}
\newcommand{\hscan}{\vname{hscan}}
\newcommand{\ID}{\vname{ID}}
\newcommand{\rec}{\vname{record}}
\newcommand{\party}{\mathcal{P}}
\newcommand{\sk}{\vname{sk}}
\newcommand{\url}[1]{#1}
\renewcommand{\Pr}[1]{\fname{Pr}\Big[#1\Big]}

\newcommand{\linebreakand}{%
  \end{@IEEEauthorhalign}
  \hfill\mbox{}\par
  \mbox{}\hfill\begin{@IEEEauthorhalign}
}

\makeatletter
\def\thanks#1{\protected@xdef\@thanks{\@thanks
        \protect\footnotetext{#1}}}
\makeatother

\begin{document}
\mainmatter

\title{Framework for a DLT Based COVID-19 Passport\\
}
\author{Sarang Chaudhari$^\dagger$ \and Michael Clear$^\ddagger$ \and Philip Bradish$^\ddagger$ \and Hitesh Tewari$^\ddagger$}
\institute{$^\dagger$Indian Institute of Technology, Delhi, $^\ddagger$Trinity College Dublin, Ireland}

\thanks{This publication has emanated from research conducted with the financial support of Science Foundation Ireland under Grant Number 13/RC/2094 (Lero).}

\maketitle

\begin{abstract}
Uniquely identifying individuals across the various networks they interact with on a daily basis remains a challenge for the digital world that we live in, and therefore the development of secure and efficient privacy preserving identity mechanisms has become an important field of research. In addition, the popularity of decentralised decision making networks such as Bitcoin has seen a huge interest in making use of distributed ledger technology to store and securely disseminate end user identity credentials. In this paper we describe a mechanism that allows one to store the COVID-19 vaccination details of individuals on a publicly readable, decentralised, immutable blockchain, and makes use of a two-factor authentication system that employs biometric cryptographic hashing techniques to generate a unique identifier for each user. Our main contribution is the employment of a provably secure input-hiding, locality-sensitive hashing algorithm over an iris extraction technique, that can be used to authenticate users and anonymously locate vaccination records on the blockchain, without leaking any personally identifiable information to the blockchain.
\end{abstract}

\section{Introduction}
Immunization is one of modern medicine’s greatest success stories. It is one of the most cost-effective public health interventions to date, averting an estimated 2 to 3 million deaths every year. An additional 1.5 million deaths could be prevented if global vaccination coverage improves \cite{WHO}. The current COVID-19 pandemic which has resulted in millions of infections worldwide \cite{JHU} has brought into sharp focus the urgent need for a ``passport" like instrument, which can be used to easily identify a user's vaccination record, travel history etc., as they traverse the globe. However, such instruments have the potential to discriminate or create bias against citizens \cite{PHELAN20201595} if they are not designed with the aim of protecting the user's identity and/or any personal information stored about them on the system.

Given the large number of potential users of such a system and the involvement of many organizations in different jurisdictions, we need to design a system that is easy to sign up to for end users, and for it to be rolled out at a rapid rate. The use of hardware devices such as smart cards or mobile phones for storing such data is going to be financially prohibitive for many users, especially those in developing countries. Past experience has shown that such ``hardware tokens" are sometimes prone to design flaws that only come to light once a large number of them are in circulation. Such flaws usually require remedial action in terms of software or hardware updates, which can prove to be very disruptive.

An alternative to the above dilemma is an \textit{online passport} mechanism. An obvious choice for the implementation of such a system is a blockchain, that provides a ``decentralized immutable ledger" which can be configured in a manner such that it can be only written to by \textit{authorised entities} (i.e. there is no requirement for a hard computation such as proof-of-work (PoW) to be carried out for monetary reward), but can be queried by anyone. However, one of the main concerns for such a system is based on: How does one preserve the privacy of user's data on a public blockchain while providing a robust mechanism to link users to their data records securely? In other words, one of the key requirements is to avoid having any personally identifiable information (PII) belonging to users stored on the blockchain.

In the subsequent sections we describe some of the key components of our system and the motivation that led us to use them. The three major components are - extraction of iris templates, a hashing mechanism to store them securely and a blockchain technology. Finally, we present a formal description of our framework which uses the aforementioned components as building blocks. However, first we will briefly discuss some related work and then briefly some preliminaries with definitions and notation that are used in the paper.

\subsection{Related Work}
There has been considerable work on biometric cryptosystems and cancellable biometrics, which aims to protect biometric data when stored for the purpose of authentication cf. \cite{journals/ejisec/RathgebU11}. Biometric hashing is one such approach that can achieve the desired property of irreversibility, albeit without salting it does not achieve unlinkability.
Research in biometric hashing for generating the same hash for different biometric templates from the same user is at an infant stage and existing work does not provide strong security assurances. Locality-sensitive hashing is the approach we explore in this paper, which has been applied to biometrics in existing work; for example a recent paper by Dang et al. \cite{Dang:2020} applies a variant of SimHash, a hash function we use in this paper, to face templates. However the technique of applying locality-sensitive hashing to a biometric template has not been employed, to the best of our knowledge, in a system such as ours.

\section{Preliminaries}
\subsection{Notation}
A quantity is said to be negligible with respect to some parameter
$\lambda$, written $\fname{negl}(\lambda)$, if it is asymptotically
bounded from above by the reciprocal of all polynomials in $\lambda$.

For a probability distribution $\mathcal{D}$, we denote by $x
\samplefrom \mathcal{D}$ the fact that $x$ is sampled according to $\mathcal{D}$. We overload the notation for a set $S$ i.e.
$y \samplefrom S$ denotes that $y$ is sampled uniformly from $S$.  Let $\mathcal{D}_0$ and $\mathcal{D}_1$ be distributions. We denote by $\mathcal{D}_0 \cind \mathcal{D}_1$ and the $\mathcal{D}_0 \sind \mathcal{D}_1$ the facts that $\mathcal{D}_0$ and $\mathcal{D}_1$ are computationally indistinguishable and statistically indistinguishable respectively. 

We use the notation $[k]$ for an integer $k$ to denote the set $\{1, \hdots, k\}$.

Vectors are written in lowercase boldface letters.

The abbreviation PPT stands for probabilistic polynomial time.

\subsection{Entropy}
The entropy of a random variable is the average ``information" conveyed by the variable's possible outcomes. A formal definition is as follows.
\begin{definition}
The entropy $H(X)$ of a discrete random variable $X$ which takes on the values $x_1, \hdots, x_n$ with respective probabilities $\Pr{X = x_1}, \hdots, \Pr{X = x_n}$ is defined as
\[
H(X) := -\sum^{n}_{i = 1}\Pr{X = x_i}\log{\Pr{X = x_i}}
\]
\end{definition}
In this paper, the logarithm is taken to be base 2, and therefore we measure the amount of entropy in bits.

\section{Iris Template Extraction}
Iris biometrics is considered one of the most reliable techniques for implementing identification systems. For the $ID$ of our system (discussed in the overview of our framework in Section~\ref{sec:framework}), we needed an algorithm that can provide us with consistent iris templates, which will have not only low intra-class variability, but also show high inter-class variability. This requirement is essential because we would expect the iris templates for the same subject to be similar, as this would then be hashed using the technique described in the section \ref{LSH}. Iris based biometric techniques have received some good attention in the last decade. One of the most successful technique was put forward by John Daugman \cite{Springer:Daugman}, but most of the current best-in-class techniques are patented and hence unavailable for an open-source use. For the purpose of writing this paper, we have used the work of Libor Masek \cite{Code:Masek} which is an open-source implementation of a reasonably reliable iris recognition technique. Users can always opt for other commercial biometric solutions when trying to deploy our work independently.

\begin{figure}[ht]
\centering
\begin{subfigure}{.45\linewidth}
  \centering
  \frame{\includegraphics[width=.9\linewidth]{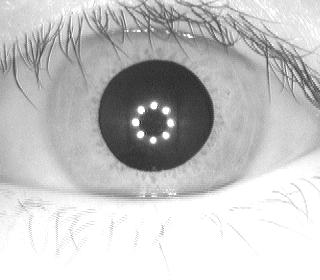}}
  \caption{}
  \label{fig:sub1}
  \vspace{0.5cm}
\end{subfigure}
\begin{subfigure}{.45\linewidth}
  \centering
  \frame{\includegraphics[width=.9\linewidth]{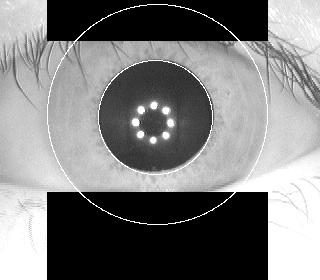}}
  \caption{}
  \label{fig:sub2}
  \vspace{0.5cm}
\end{subfigure}
\begin{subfigure}{0.96\linewidth}
  \centering
  \frame{\includegraphics[width=.9\linewidth]{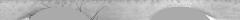}}
  \caption{}
  \label{fig:sub3}
  \vspace{0.5cm}
\end{subfigure}
\begin{subfigure}{0.96\linewidth}
  \centering
  \frame{\includegraphics[width=.9\linewidth]{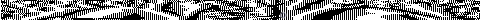}}
\end{subfigure}
\begin{subfigure}{0.96\linewidth}
  \centering
  \frame{\includegraphics[width=.9\linewidth]{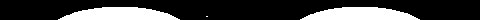}}
  \caption{}
  \label{fig:sub5}
\end{subfigure}
\caption{Masek's Iris Template Extraction Algorithm}
\label{fig:iris}
\end{figure}

Masek's technique works on grey-scale eye images, which are processed in order to extract the binary template. First the segmentation algorithm, based on a Hough Transform is used to localise the iris and pupil regions and also isolate the eyelid, eyelash and reflections as shown in Figures \ref{fig:sub1} and \ref{fig:sub2}. The segmented iris region is then normalised i.e, unwrapped into a rectangular block of constant polar dimensions as shown in Figure \ref{fig:sub3}. The iris features are extracted from the normalised image by one-dimensional Log-Gabor filters to produce a bit-wise iris template and mask as shown in Figure \ref{fig:sub5}. We denote the complete algorithm by $\aname{Iris.ExtractFeatureVector}$ which takes a scanned image as input and outputs a binary feature vector $\vect{fv} \in \{0, 1\}^n$ (this algorithm is called upon by our framework in Section~\ref{sec:framework}). This data is then used for matching, where the Hamming distance is used as the matching metric. We have used CASIA-Iris-Interval database \cite{CASIA} in Figure \ref{fig:iris} and for some preliminary testing.

\begin{table}[ht]
\begin{tabularx}{\linewidth}{ 
  || >{\centering\arraybackslash}X 
  || >{\centering\arraybackslash}X 
  | >{\centering\arraybackslash}X || }
\hline
Threshold & FAR & FRR \\ [0.5ex]
\hline
\hline
0.20 & 0.000 & 74.046 \\
\hline
0.25 & 0.000 & 45.802 \\
\hline
0.30 & 0.000 & 25.191 \\
\hline
0.35 & 0.000 & 4.580 \\
\hline
0.40 & 0.005 & 0.000 \\
\hline
0.45 & 7.599 & 0.000 \\
\hline
0.50 & 99.499 & 0.000 \\
\hline
\end{tabularx}
\caption {FAR and FRR for the CASIA-a Data Set}
\centering
\label{table:table_libor_res}
\end{table}
\addtolength{\topmargin}{0.03 in}

Table \ref{table:table_libor_res} shows the performance of Masek's technique as reported by him in his original thesis \cite{Thesis:Masek}. This algorithm performs quite well for a threshold of 0.4 where the false acceptance rate (FAR) is 0.005 and the false rejection rate (FRR) is 0. These values are used when we present our results to compare the performance of our technique when a biometric template is first hashed and then the hamming distance is measured to calculate the FAR and FRR, as opposed to directly measuring the hamming distances in the original biometric templates. One would assume the performance of our work to get better with the increase in efficiency of extracting consistent biometric templates by other methods.

As aforementioned, we rely on the open-source MATLAB code by Libor Masek. For each input image, the algorithm produces a binary template which contains the iris information, and a corresponding noise mask which corresponds to corrupt areas within the iris pattern, and marks bits in the template as corrupt. These extracted iris templates and their corresponding masks are $20 \times 480$ binary matrices each. In the original work, only those bits in the iris pattern that correspond to ‘0’ bits in the noise masks of both iris patterns were used in the calculation of Hamming distance. A combined mask is then calculated and both the templates are masked with it. Finally, the algorithm calculates the bitwise XOR i.e. distance between the masked templates. The steps given below provides an overview for the algorithm used by Libor: \\ \newline
Extraction:
\[
(template_1, \, mask_1) = createiristemplate(\mathbf{image_1})
\]
\[
(template_2, \, mask_2) = createiristemplate(\mathbf{image_2})
\]
\newline
Matching:
\[
c\_mask = mask_1 \land mask_2
\]
\[
masked\_template_1 = template_1 \land (\lnot \, c\_mask)
\]
\[
masked\_template_2 = template_2 \land (\lnot \, c\_mask)
\]
\[
distance = masked\_template_1 \oplus masked\_template_2 
\]

There are two major issues that we need to deal with before being able to use these templates in our system, specifically, template masking and conversion to linear vector.

\subsection{Template Masking}\label{sec:tm}
The above matching technique requires one to have 2 pairs of iris patterns and their corresponding masks to calculate the Hamming distance. However for the application we are targeting, at any time during verification, the system would have to match the extracted template of an individual (i.e. $template$ and $mask$) against a hashed template stored on the blockchain. This means that we cannot incorporate the above matching algorithm into our system. We have two choices to mitigate this problem and obtain the masked\_template for the remaining steps:
\begin{enumerate}
  \item We can calculate the $masked\_template$ independently for each sample i.e.
    \[
    masked\_template = template \land (\lnot \, mask)
    \]
  An underlying assumption for this method is that the masks for an individual would be approximately the same in every sample. This assumption is not too far-fetched as was clear from the preliminary analysis of our database. We will refer to these as $type_1$ templates.
  
  \item We can maintain a $global\_mask$ which can be the defined as
    \[
    global\_mask = mask_1 \land \hdots \land mask_l
    \]
    for all extracted $mask_i$ of all $image_i$ belonging to the training database. And at the time of verification, we generate the masked\_template as
    \[
    masked\_template = template \land (\lnot \, global\_mask)
    \]
    This method has some added difficulty in finding the $global\_mask$ and it also discards more data from the iris patterns as opposed to directly using the respective $mask_i$ for each $template_i$. But it helps in maintaining a consistency among the $masked\_templates$. We will refer to these as $type_2$ templates.
\end{enumerate}
In a follow-up to this paper, we will use and provide results for templates of both types based on experiments we are conducting at the time of writing.

\subsection{Conversion to Linear Vector}\label{conversion_to_linear_vector}
For our use case, we need a one-dimensional input stream which can be fed into the hashing algorithm discussed in the subsequent sections. For converting those $masked\_template$ matrices into linear feature vectors, we have two naive choices of concatenating either the row vectors or the column vectors. Before deciding the type of conversion, let us look at an important key factor, which is \textit{rotational inconsistencies} in the iris templates.
\par
Rotational inconsistencies are introduced due to rotations of the camera, head tilts and rotations of the eye within the eye socket. The normalisation process does not compensate these. In order to account for rotational inconsistencies, when the Hamming distance of two templates is calculated, one template is shifted left and right bit-wise, and a number of Hamming distance values are calculated from successive shifts. This bit-wise shifting in the horizontal direction corresponds to the rotation of the original iris region. This method was suggested by Daugman \cite{Springer:Daugman}, and corrects misalignment in the normalised iris pattern caused by rotational differences during imaging. From the calculated Hamming distance values, only the lowest is taken, since this corresponds to the best match between two templates. Due to this, column-wise conversion seems like the most logical choice as this would allow us to easily rotate the binary linear feature vectors. Shifting the linear vector by 20 bits will correspond to shifting the iris template once (recall that the dimension of iris templates is $20 \times 480$).

\subsection{Wrap-up}
Putting it all together, we define the steps of our algorithm $\aname{Iris.ExtractFeatureVector}$ that we call upon later.
\begin{itemize}
    \item $\mathbf{Iris.ExtractFeatureVector}(\vname{image})$:
    \begin{itemize}
        \item $(template, mask) = createiristemplate(\vname{image})$ where $createiristemplate$ is Masek's open source algorithm.
        \item Obtain $masked\_template$ (either $type_1$ or $type_2$ as defined in Section~\ref{sec:tm}).
        \item Convert $masked\_template$ to linear vector as in Section~\ref{conversion_to_linear_vector}.
        \item Output binary linear vector $\vect{\vname{fv}} \in \{0, 1\}^n$
    \end{itemize}
\end{itemize}
Note that the parameter $n$ is a global system parameter measuring the length of the binary feature vectors outputted by $\aname{Iris.ExtractFeatureVector}$.

\section{Locality-sensitive Hashing}\label{LSH}
To preserve the privacy of individuals on the blockchain, the biometric data has to be encrypted before being written to the ledger. Hashing is a good \textit{alternative} to achieve this, but techniques such as SHA-256 and SHA-3 cannot be used, since the biometric templates that we extracted above can show differences across various scans for the same individual. Hence using those hash functions would produce completely different hashes. Therefore, we seek a hash function that generates ``similar" hashes for similar biometric templates. This prompts us to explore Locality-Sensitive Hashing (LSH), which has exactly this property. Various LSH techniques have been researched to identify whether files (i.e. byte streams) are similar based on their hashes. TLSH is a well-known LSH function that exhibits high performance and matching accuracy but, does not provide a sufficient degree of security for our application. Below we assess another type of LSH function, which does not have the same runtime performance as TLSH, but as we shall see, exhibits provable security for our application and therefore is a good choice for adoption in our framework.

\subsection{Input Hiding}\label{sec:ih}
In the cryptographic definition of one-way functions, it is required that it is hard to find \emph{any} preimage of the function. However, we can relax our requirements for many applications because it does not matter if for example a random preimage can be computed as long as it is hard to learn information about the specific preimage that was used to compute the hash. In this section, we introduce a property that captures this idea, a notion we call \emph{input hiding}.

\emph{Input hiding} means that if we choose some preimage $x$ and give the hash $h = H(x)$ to an adversary, it is either computationally hard or information-theoretically impossible for the adversary to learn $x$ or any partial information about $x$. This is captured in the following formal definition.
\begin{definition}
A hash function family $\mathcal{H}$ with domain $X := \{0, 1\}^n$ and range $\mathcal{Y} := \{0, 1\}^m$ is said to be input hiding if for all randomly chosen hash functions $H \samplefrom \mathcal{H}$, all $i \in [n]$, all randomly chosen inputs $x \samplefrom X$  and all PPT adversaries $\mathcal{A}$ it holds that
\begin{align*}
\Big|\Pr{x_i = 0 \; \land \; \mathcal{A}(i, H(x)) \rightarrow 1} -& \\
\Pr{x_i = 1 \; \land \; \mathcal{A}(i, H(x)) \rightarrow 1}\Big| & \leq \fname{negl}(\lambda)
\end{align*}
where $\lambda$ is the security parameter.
\end{definition}

\subsection{Our Variant of SimHash}\label{sec:randproj}
Random projection hashing, proposed by Charikar \cite{ACM:Charikar}, preserves the cosine distance between two vectors in the output of the hash, such that two hashes are probabilistically similar depending on the cosine distance between the two preimage vectors. This hash function is called SimHash. We describe a slight variant of SimHash here, which we call $\vname{S3Hash}$. In our variant, the random vectors that are used are sampled from the finite field of $\mathbb{F}_3 = \{-1, 0, 1\}$. Suppose we choose a hash length of $m$ bits. Now for our purposes, the input vectors to the hash are binary vectors in $\{0, 1\}^n$ for some $n$. First we choose $m$ random vectors $\vect{r}_i \samplefrom \{-1, 0, 1\}^n$ for $i \in \{1, \hdots, m\}$. Let $R = \{\vect{r}_i\}_{i \in \{1, \hdots, m\}}$ be the set of these random vectors. The hash function $\fname{S3Hash}_R : \{0, 1\}^n \to \{0, 1\}^m$ is thus defined as:
\begin{equation}
\fname{S3Hash}_R(\vect{x}) = (\mathsf{sgn}(\dotprod{\vect{x}}{\vect{r}_1}), \hdots, \mathsf{sgn}(\dotprod{\vect{x}}{\vect{r}_m}))
\end{equation}
where $\mathsf{sgn} : \mathbb{Z} \to \{0, 1\}$ returns $0$ if its integer argument is negative and returns $1$ otherwise. Note that the notation $\dotprod{\cdot}{\cdot}$ denotes the inner product between the two specified vectors. Let $\vect{x}_1, \vect{x}_2 \in \{0, 1\}^n$ be two input vectors. It holds for all $i \in \{1, \hdots, m\}$ that $\mathsf{Pr}[h^{(1)}_i = h^{(2)}_i] = 1 - \frac{\theta(\vect{x}_1, \vect{x}_2)}{\pi}$ where $\vect{h}^{(1)} = \fname{S3Hash}_R(\vect{x}_1)$, $\vect{h}^{(2)} = \fname{S3Hash}_R(\vect{x}_2)$ and $\theta(\vect{x}_1, \vect{x}_2)$ is the angle between $\vect{x}_1$ and $\vect{x}_2$. Therefore the similarity of the inputs is preserved in the similarity of the hashes.

An important question is: Is this hash function suitable for our application? The answer is in the affirmative because it can be proved that the function information-theoretically obeys a property we call \emph{input-hiding} that we defined in Section~\ref{sec:ih}. We recall that this property  means that if we choose some binary vector $\vect{x} \in \{0, 1\}^n$ and give the hash $\vect{h} = \fname{S3Hash}_R(\vect{x})$ to an adversary, it is either computationally hard or information-theoretically impossible for the adversary to learn $\vect{x}$ or any partial information about $\vect{x}$. This property is sufficient in our application since we only have to ensure that no information is leaked about the user's iris template. We now prove that our variant locality-sensitive hash function $\vname{S3Hash}$ is information-theoretically input hiding.
\vspace{6pt}

\begin{theorem}
Let $X$ denote the random variable corresponding to the domain of the hash function. If $H(X) \geq m + \lambda$ then S3Hash is information-theoretically input hiding where $\lambda$ is the security parameter and $H(X)$ is the entropy of $X$.
\end{theorem}
\vspace{6pt}
\begin{proof}
The random vectors in $R$ can be thought of as vectors of coefficients corresponding to a set of $m$ linear equation in $n$ unknowns on the left hand side and on the right hand side we have the $m$ elements, one for each equation, which are components of the hash i.e. $(h_1, \hdots, h_m)$. Now the inner product is evaluated over the integers and the $\fname{sgn}$ function maps an integer to an element of $\{0, 1\}$ depending on its sign. The random vectors are chosen to be ternary. Suppose we choose a finite field $\mathbb{F}_p$ where $p \geq 2(m + 1)$ is a prime. Since there will be no overflow when evaluating the inner product in this field, a solution in this field is also a solution over the integers. We are interested only in the binary solutions. Because $m < n$, the system is underdetermined. Since there are $n - m$ degrees of freedom in a solution, it follows that there are $2^{n - m}$ binary solutions and each one is equally likely. Now let $r$ denote the redundancy of the input space i.e. $r = n - H(X)$. The fraction of the $2^{n - m}$ solutions that are \emph{valid} inputs is $2^{n - m - r}$. If $2^{n - m - r} > 2^{\lambda}$, then the probability of an adversary choosing the ``correct" preimage is negligible in the security parameter $\lambda$. For this condition to hold, it is required that $n - m - r > \lambda$ (recall that $r = n - H(X)$), which follows if $H(X) \geq m + \lambda$ as hypothesized in the statement of the theorem. It follows that information-theoretically an unbounded adversary has a negligible advantage in the input hiding definition.
\end{proof}

Our initial estimates suggest that the entropy of the distribution of binary feature vectors outputted by $\aname{Iris.ExtractFeatureVector}$ is greater than $m + \lambda$ for parameter choices such as $m = 256$ and $\lambda = 128$. A more thorough analysis however is deferred to future work.

\subsection{Evaluation}
We have ran experiments with $\aname{S3Hash}$ applied to feature vectors obtained using our $\aname{Iris.ExtractFeatureVector}$ algorithm. The distance measure we use is the hamming distance. The results of these experiments are shown in Table~\ref{table:rp_results}. Results for a threshold of 0.3 in particular indicates that our approach shows promise. We hope to make further improvements in future work.
\begin{table}[ht]
\begin{tabularx}{\linewidth}{ 
  || >{\centering\arraybackslash}X 
  || >{\centering\arraybackslash}X 
  | >{\centering\arraybackslash}X || }
\hline
Threshold & FAR & FRR \\ [0.5ex]
\hline
\hline
0.25       &          3.99     &     64.26 \\
\hline
0.26       &          6.35     &     57.35 \\
\hline
0.27       &          9.49     &     49.63 \\
\hline
0.28       &         13.92     &    43.79 \\
\hline
0.29       &         19.60     &    36.47 \\
\hline
0.3        &          26.23    &     30.79 \\
\hline
0.31       &         34.01     &    25.10 \\
\hline
0.32       &         42.30     &    19.33 \\
\hline
0.33       &         50.91     &    16.00 \\
\hline
0.34       &         59.04     &    11.94 \\
\hline
0.35       &         67.05     &     8.53 \\
\hline
\end{tabularx}
\caption {FAR \& FRR for the CASIA-Iris-Interval Data Set}
\centering
\label{table:rp_results}
\end{table}

\section{Blockchain}
A blockchain is used in the system for immutable storage of individuals' vaccination records. The blockchain we employ is a \textit{permissioned} ledger to which blocks can only be added by authorized entities or persons such as hospitals, primary health care centers, clinicians etc. Such entities have to obtain a public-key certificate from a trusted third party and store it on the blockchain as a transaction before they are allowed to add blocks to the ledger. The opportunity to add a new block is controlled in a \emph{round robin} fashion, thereby eliminating the need to perform a computationally intensive PoW process. Any transactions that are broadcast to the P2P network are signed by the entity that created the transaction, and can be verified by all other nodes by downloading the public key of the signer from the ledger itself. An example of distributed ledger technology that fulfills the above requirements is MultiChain \cite{DLT:Multichain}.

\subsection{Interface}
We now describe an abstract interface for the permissioned blockchain that captures the functionality we need. Consider a set of parties $\hat{\mathbb{P}}$. A subset of parties $\mathbb{P} \subset \hat{\mathbb{P}}$ are authorized to write to the blockchain. Each party $\mathcal{P} \in \mathbb{P}$ has a secret key $\sk_\mathcal{P}$ which it uses to authenticate itself and gain permission to write to the blockchain. How a party acquires authorization is beyond the scope of this paper. For our purposes, the permissioned blockchain consists of the following algorithms:
\begin{itemize}
    \item $\aname{Blockchain.Broadcast}(\mathcal{P}, \sk_\mathcal{P}, \vname{tx})$: On input a party identifier $\mathcal{P}$ that identifies the sending party, a secret key $\sk_\mathcal{P}$ for party $\mathcal{P}$ and a transaction $\vname{tx}$ (whose form is described below), then broadcast the transaction $\vname{tx}$ to the peer-to-peer network for inclusion in the next block. The transaction will be included iff $\mathcal{P} \in \mathbb{P}$.
\item
$\aname{Blockchain.AnonBroadcast}(\sk_\mathcal{P}, \vname{tx}):$ On input a secret key $\sk_\party$ for a party $\party$ and a transaction $\vname{tx}$, then anonymously broadcast the transaction $\vname{tx}$ to the peer-to-peer network for inclusion in the next block. The transaction will be included iff $\mathcal{P} \in \mathbb{P}$.
\item
$\aname{Blockchain.GetNumBlocks}()$: Return the total number of blocks currently in the blockchain.

\item
$\aname{Blockchain.RetrieveBlock}(\vname{blockNo})$: Retrieve and return the block at index $\vname{blockNo}$, which is a non-negative integer between 0 and $\aname{Blockchain.GetNumBlocks}() - 1$.
\end{itemize}
A transaction has the form $(\vname{type}, \vname{payload}, \vname{party}, \vname{signature})$. A transaction in an anonymous broadcast is of the form $(\vname{type}, \vname{payload}, \bot, \bot)$. The payload of a transaction is interpreted and parsed depending on its type. In our application, there are two permissible types: 'rec' (a record transaction which consists of a pair (ID, record)) and 'hscan' (biometric hash transaction which consists of a hash of an iris feature vector). This will become clear from context in our formal description of our framework in the next section which makes use of the above interface as a building block. The final point is that a block is a pair $(\vname{hash}, \vname{transactions})$ consisting of the hash of the block and a set of transactions $\{\vname{tx}_i\}_{i \in [\ell]}$.

\section{Our Framework}\label{sec:framework}

\subsection{Overview}
In this section we provide a formal description of our proposed framework which makes use of the building blocks presented in the previous sections. Our proposed system utilises a two-factor authentication mechanism to uniquely identify an individual on the blockchain. The parameters required to recreate an identifier are based on information that ``one knows" and biometric information that ``one possess".

\begin{figure}[htbp]
\centering
\includegraphics[width=4in]{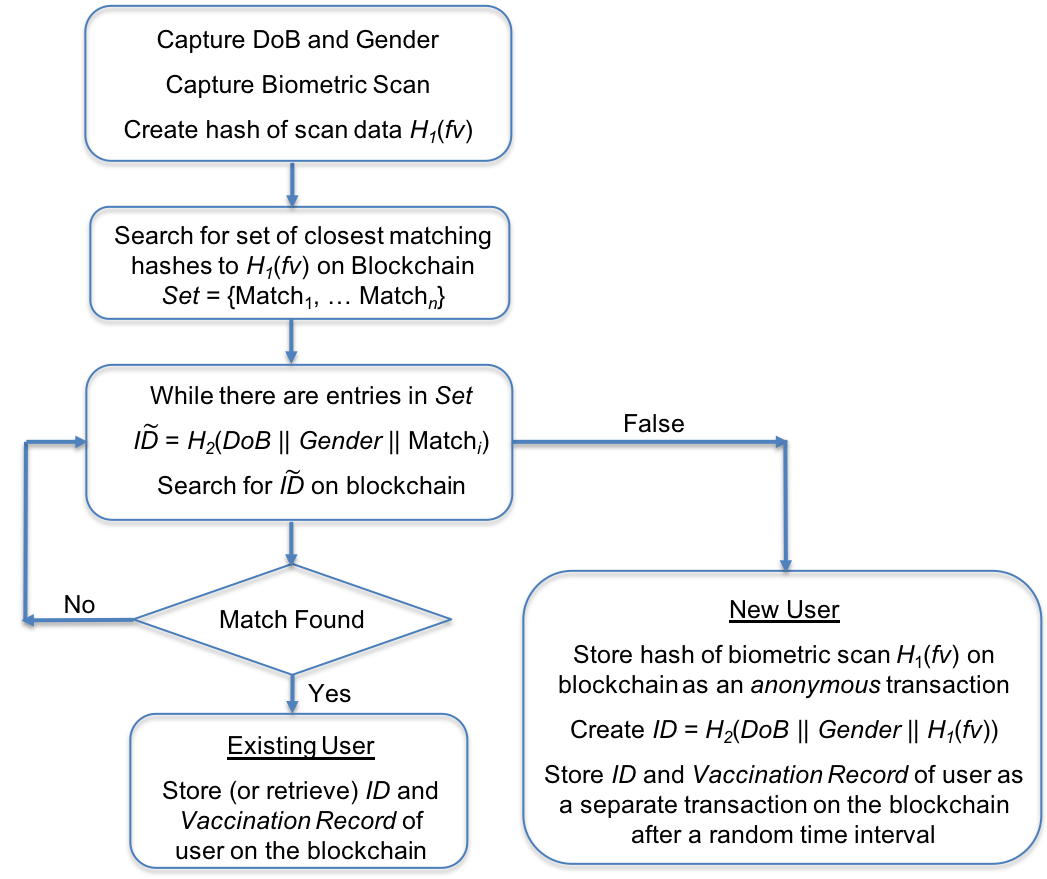}
\caption{Algorithm Workflow}
\label{fig_alg}
\end{figure}

Figure \ref{fig_alg} describes the overall algorithm that we employ in our proposed system. When a user presents themselves to an entity or organisation participating in the system, they are asked for their \textit{DoB}(dd/mm/yyyy) and \textit{Gender}(male/female/other). In addition, the organization captures a number of scans of the user's iris, and creates a hash ${H_1}(\vect{fv})$ from the feature vector extracted from the ``best" biometric scan data. Our system can combine the user's \textit{DoB} and \textit{Gender} with ${H_1}(\vect{fv})$ to generate a unique 256-bit identifier ($ID$) for the user:

\begin{equation}
ID = {H_2}(DoB~||~Gender~||~{H_1}(\vect{fv}))
\label{eqn1}
\end{equation}

The algorithm tries to match the calculated hash $H_1(\vect{fv})$ with existing ``anonymous" hashes that are stored on the blockchain. It may get back a set of hashes that are somewhat ``close" to the calculated hash. In that case the algorithm concatenates each returned hash ($Match_i$) with the user's \textit{DoB} and \textit{Gender} to produce $\widetilde{ID}$. It then tries to match $\widetilde{ID}$ with an $ID$ in a vaccination record transaction on the blockchain.

If a match is found then the user is already registered on the system and has at least one vaccination record. At this point we may just wish to retrieve the user's records or add an additional record, e.g. when a booster dose has been administered to the user. However if we go through the set of returned matches and cannot match $\widetilde{ID}$ to an existing $ID$ in a vaccination record on the blockchain, i.e. this is the first time the user is presenting to the service, then we store the iris scan hash data $H_1(\vect{fv})$ as an anonymous record on the blockchain, and subsequently the $ID$ and COVID-19 vaccination details for the user as a separate transaction. In each case the transaction is broadcast at a \textit{random interval} on the blockchain peer-to-peer (P2P) network for it to be verified by other nodes in the system, and eventually added to a block on the blockchain. Uploading the two transactions belonging to a user at random intervals ensures that the transactions are stored on separate blocks on the blockchain, and an attacker is not easily able to identify the relationship between the two.

\begin{figure}[htbp]
\centering
\includegraphics[width=4.5in]{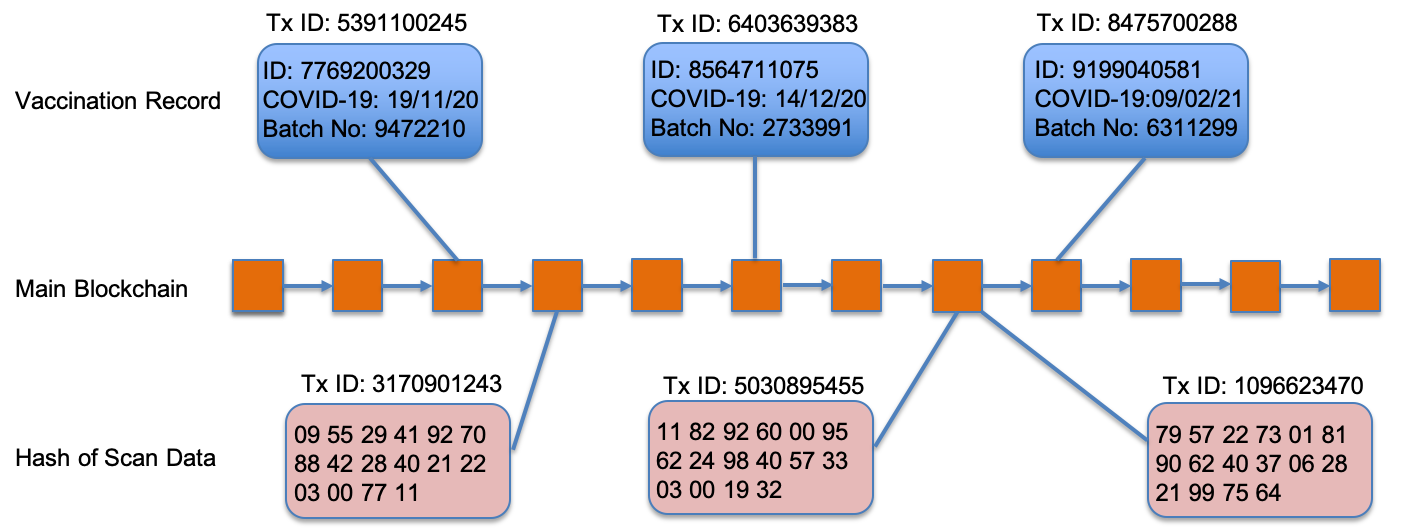}
\caption{Blockchain Structure}
\label{fig_blockchain}
\end{figure}

Figure \ref{fig_blockchain} shows a blockchain in which there are three anonymous transactions (i.e. Hash of Scan Data) and three COVID-19 vaccination record transactions stored on the blockchain pertaining to different users. The reader is referred to Section \ref{LSH} for more details on how the hash is calculated in our system. Note that the storage of the anonymous hash data has to be carried out only \textit{once} per registered user in the system.

\addtolength{\topmargin}{0.02 in}

\subsection{Formal Description}
We present a formal description of our framework in Figure~\ref{fig:framework} and Figure~\ref{fig:framework2}. Note that the algorithms described in these figures are intended to formally describe the fundamental desired functionality of our framework and are so described for ease of exposition and clarity; in particular, they are naive and non-optimized, specifically not leveraging more efficient data structures as would a real-world implementation.

Let $\mathcal{H}$ be a family of collision-resistant hash functions. The algorithms in Figure~\ref{fig:framework} are stateful (local variables that contain retrieved information from the blockchain are shared and accessible to all algorithms). Furthermore, the parameters tuple $\params$ generated in $\aname{Setup}$ is an implicit argument to all other algorithms.

\begin{figure}[!ht]
\begin{small}
\begin{center}
\hspace{-19pt}
\begin{tabular}{l}
\begin{minipage}{3in}\vspace{6pt}
\begin{tabbing}
123\=12\=12\=12\=12\=\kill
\textbf{Algorithm} $\aname{Setup}(1^{\lambda})$ \\
        \> $\vect{r}_i \samplefrom \{-1, 0, 1\}^n$ for $i \in \{1, \hdots, m\}$ for $i \in [m]$. \\
        \> $R \gets \{\vect{r}_1, \hdots, \vect{r}_m\}$ \\
        \> $H_1 \samplefrom \mathcal{H}$ \\
        \> $H_2 \gets \aname{S3Hash}_R$ \\
        \> $\nblocks \gets 0$ \\
        \> $\ids \gets \emptyset$\\
        \> $\records \gets \emptyset$\\
        \> $\hscans \gets \emptyset$\\
        \> $\aname{Sync}()$\\
        \> Return $\params := (H_1, H_2)$
\end{tabbing}
\end{minipage} \\
\begin{minipage}{3in}\vspace{6pt}
\begin{tabbing}
123\=12\=12\=12\=12\=\kill
\textbf{Algorithm} $\aname{AddRecord}(\party, \sk_\party, \vname{dob}, \vname{gender}, \vname{scan}, \rec)$ \\
        \> $\ID \gets \aname{Authenticate}(\vname{dob}, \vname{gender}, \vname{scan})$\\
        \> If $\ID = \bot$:\\
        \>\> $\ID \gets \aname{Enroll}(\party, \sk_\party, \vname{dob}, \vname{gender}, \vname{scan}, \rec)$\\
        \>\> Return\\
        \> $\vname{payload} \gets (\ID, \rec)$\\
        \> $\sigma \gets \aname{Sign}(\sk_\party, \vname{payload})$\\
        \> $\vname{tx} \gets (\text{'rec'}, \vname{payload}, \party, \sigma)$\\
        \> $\aname{Blockchain.Broadcast}(\party, \sk_\party, \vname{tx})$\\
        \> $\records \gets \records \cup \{(\ID, \rec)\}$\\
\end{tabbing}
\end{minipage} \\
\begin{minipage}{3in}\vspace{6pt}
\begin{tabbing}
123\=12\=12\=12\=12\=\kill
\textbf{Algorithm} $\aname{FetchRecords(\vname{dob}, \vname{gender}, \vname{scan}})$ \\
        \> $\ID \gets \aname{Authenticate}(\vname{dob}, \vname{gender}, \vname{scan})$\\
        \> If $\ID = \bot$:\\
        \>\> Return $\emptyset$ \\
        \> $\vname{results} \gets \{\rec : (\ID, \rec) \in \records\}$\\
        \> Return $\vname{results}$
\end{tabbing}
\end{minipage} \\
\end{tabular}
\end{center}
\caption{{\small
Our Framework for a COVID-19 Passport
}}
\label{fig:framework}
\end{small}
\end{figure}

The algorithms invoked by the algorithms in Figure~\ref{fig:framework} can be found in Figure~\ref{fig:framework2}.

\begin{figure}[!ht]
\begin{small}
\begin{center}
\hspace{-19pt}
\begin{tabular}{l}
\begin{minipage}{3in}\vspace{6pt}
\begin{tabbing}
123\=12\=12\=12\=12\=\kill
\textbf{Algorithm} $\aname{Authenticate}(\vname{dob}, \vname{gender}, \vname{scan})$ \\
        \> $\aname{Sync}()$\\
        \> $\vect{fv} \gets \aname{Iris.ExtractFeatureVector}(\vname{scan})$\\
        \> $\hscan \gets H_2(\vect{fv})$\\
        \> $\ID \gets H_1(\vname{dob} \concat \vname{gender} \concat \hscan)$\\
        \> If $\ID \in \ids$:\\
        \>\> Return $\ID$\\
        \> For each $h \in \hscans$:\\
        \>\> $d \gets \aname{Dist}(\hscan, h)$\\
        \>\> If $d < \vname{THRESHOLD}$:\\
        \>\>\> $\widetilde{\ID} \gets H_1(\vname{dob} \concat \vname{gender} \concat h)$\\
        \>\>\> If $\widetilde{ID} \in \ids$:\\
        \>\>\>\> Return $\widetilde{\ID}$\\
        \> Return $\bot$
\end{tabbing}
\end{minipage} \\
\begin{minipage}{3in}\vspace{6pt}
\begin{tabbing}
123\=12\=12\=12\=12\=\kill
\textbf{Algorithm} $\aname{Enroll}(\party, \sk_\party, \vname{dob}, \vname{gender}, \vname{scan}, \vname{initRecord})$ \\
        \> $\aname{Sync}()$\\
        \> $\vect{fv} \gets \aname{Iris.ExtractFeatureVector}(\vname{scan})$\\
        \> $\hscan \gets H_2(\vect{fv})$\\
        \> $\ID \gets H_1(\vname{dob} \concat \vname{gender} \concat \hscan)$\\
        \> $\vname{payload} \gets (\ID, \vname{initRecord})$\\
        \> $\sigma \gets \aname{Sign}(\sk_\party, \vname{payload})$\\
        \> $\vname{tx} \gets (\text{'rec'}, \vname{payload}, \party, \sigma)$\\
        \> $\aname{Blockchain.Broadcast}(\party, \sk_\party, \vname{tx})$\\
        \> $t \samplefrom \{1, \hdots, 100\}$\\
        \> $\vname{tx}' \gets (\text{'hscan'}, \hscan)$\\
        \> Queue execution of $\aname{Blockchain.AnonBroadcast}(\sk_\party, \vname{tx}')$\\ \> after time $t$\\
        \> $\ids \gets \ids \cup \{\ID\}$\\
        \> $\hscans \gets \hscans \cup \{\hscan\}$\\
        \> $\records \gets \records \cup \{(\ID, \vname{initRecord})\}$\\
        \> Return $\ID$
\end{tabbing}
\end{minipage} \\
\begin{minipage}{3in}\vspace{6pt}
\begin{tabbing}
123\=12\=12\=12\=12\=\kill
\textbf{Algorithm} $\aname{Sync}()$ \\
        \> $\vname{newNumBlocks} \gets \aname{Blockchain.GetNumBlocks}()$\\
        \> If $\vname{newNumBlocks} > \nblocks$:\\
        \>\> For $\nblocks \leq i < \vname{newNumBlocks}$:\\
        \>\>\> $\vname{block} \gets \aname{Blockchain.RetrieveBlock}(i)$\\
        \>\>\> $(\vname{hash}, \vname{transactions}) \gets \vname{block}$\\
        \>\>\> For each $\vname{tx} \in \vname{transactions}$:\\
        \>\>\>\> $(\vname{type}, \vname{payload}, \cdot, \cdot) \gets \vname{tx}$\\
        \>\>\>\> If $\vname{type} = \text{'rec'}$:\\
        \>\>\>\>\> $(\ID, \rec) \gets \vname{payload}$\\
        \>\>\>\>\> $\ids \gets \ids \cup \{\ID\}$\\
        \>\>\>\>\> $\records \gets \records \cup \{(\ID, \rec)\}$\\
        \>\>\>\> Else if $\vname{type} = \text{'hscan'}$:\\
        \>\>\>\>\> $\hscan \gets \vname{payload}$\\
        \>\>\>\>\> $\hscans \gets \hscans \cup \{\hscan\}$\\
        \> $\nblocks \gets \vname{newNumBlocks}$
\end{tabbing}
\end{minipage} \\
\end{tabular}
\end{center}
\caption{{\small
Additional Algorithms Used By Our Framework
}}
\label{fig:framework2}
\end{small}
\end{figure}

\section{Conclusions and Future Work}
In this paper we have detailed a framework to build a global vaccination passport using a distributed ledger. The main contribution of our work is to combine a Locality-sensitive hashing mechanism with a blockchain to store the vaccination records of users. A variant of the SimHash LSH function is used to derive an identifier that leaks no personal information about an individual. The only way to extract a user's record from the blockchain is by the user presenting themselves in person to an authorised entity, and providing an iris scan along with other personal data in order to derive the correct user identifier. However, our research has raised many additional challenges and research questions whose resolution require further investigation and experimentation, intended for future work. First and foremost, we need to improve the accuracy of the $\aname{Iris.ExtractFeatureVector}$ algorithm (e.g: deciding whether to use $type_1$ or $type_2$ template masking) and to accurately compute the entropy of the feature vectors.

Furthermore, our variant of the $\aname{SimHash}$ algorithm, referred to in this paper as $\aname{S3Hash}$, requires further analysis and evaluation, especially with respect to the domain of the random vectors $\vect{r}_i$, which are restricted to be ternary in this paper. Additionally, we must choose a suitable blockchain. Finally, the overall protocol would benefit from a thorough security analysis where not just privacy but other security properties are tested.

The blockchain based mechanism that we have proposed can also be used as a \textit{generalised healthcare} management system \cite{IEEE:Hanley, IEEE:Tewari} with the actual data being stored off-chain for the purpose of efficiency. Once a user's identifier has been recreated it can be used to pull all records associated with the user, thereby retrieving their full medical history. We are in the middle of developing a prototype implementation of the system and hope to present the results of our evaluation in a follow-on paper. At some time in the future, we hope to trial the system in the field, with the hope of rolling it out on a larger scale. Our implementation will be made open source.

\bibliographystyle{abbrv}
\bibliography{bibliography/main}

\end{document}